# Synchronized Rebuilds: The Hidden Cost of Concurrent Inserts in Learned Indexes

Emmilly Immaculate Namuganga
MPI-SWS
Saarbrücken, Germany
enamugan@mpi-sws.org

Laurent Bindschaedler
MPI-SWS
Saarbrücken, Germany
bindsch@mpi-sws.org

## Abstract

Nine years after learned indexes were proposed, their advantage on read-mostly data is well established, and a standard benchmark rates their best concurrent version as almost ready for use. Across more than 4000 runs of 13 indexes, we find that each design pays for concurrent inserts in a different way. Gapped-array indexes rebuild their nodes in synchronized bursts, at positions that follow from three constants in the source. Model-placement indexes use three to four times the memory, buffered indexes take two to three times as long. Three of five concurrent learned indexes lost keys or stopped making progress. Standard benchmarks miss both the bursts and the lost keys. The bursts cost far more than in B-trees because each rebuild copies a node a thousand times larger than a page and, under concurrency, blocks every operation that reaches the node. The typical B-tree fix of randomizing node density only spreads the bursts out, while smaller nodes trade them for longer run times at scale. Two standard systems techniques remove them: a buffer that exists only while a node is rebuilt, and background rebuilds. We show that on 400M real keys, these techniques reduce the number of inserts whose latency exceeds 100 µs by two orders of magnitude at one thread, and by one at sixteen when spare cores run the background threads, at a third to half the run time of the buffered designs.

## 1 Introduction

Learned indexes replace the comparisons in a search tree with models that predict where a key is stored [11]. Nine years after they were proposed, their advantage on read-mostly data is well established: over keys a model fits well, they match a B-tree's lookup latency in much less memory. Updatable versions absorb inserts by leaving free slots in their data nodes, by placing keys at predicted slots, or by buffering new keys. GRE finds the concurrent version of ALEX almost ready [33], while Sun et al. find learned indexes behind classical ones under concurrency [28].

We measure where these indexes stand once they must absorb inserts concurrently and at scale. We benchmarked 13 index implementations, 9 of them learned, in more than 4000 runs on five real datasets and two synthetic ones of up to 800M keys, with 1 to 32 threads and workloads ranging from read-heavy to insert-only. We find that under concurrent inserts the cost comes from how these indexes restructure. Each design pays for inserts somewhere the standard benchmarks do not look. Gapped-array indexes rebuild their nodes in synchronized bursts. Model-placement indexes avoid the bursts but use three to four times the memory, while buffered indexes avoid them at two to three times the run time. Under stress, three of the five concurrent learned indexes we ran lost keys or stopped making progress, two of them with our substitute for a proprietary math routine. Standard benchmarks miss these losses because they never check afterward that the inserted keys are still there.

We examine the bursts in most depth, because they show most directly how restructuring becomes insert latency. A gapped-array index rebuilds a data node into a larger one once the fraction of its slots in use, its *density*, crosses a threshold. Bulk loading, which is how these indexes are built, gives every node the same density. As Figure 1a shows, the nodes then reach the threshold together, at points in the insert stream that three constants in the index's source predict. B-trees show the same effect, known as waves of misery [8]. In learned indexes it costs far more, both in insert latency and in run time, because a rebuild copies tens of thousands of entries where a B-tree split copies a page of tens. Under concurrency, it also locks its node, so every other insert and lookup that reaches the node restarts, as Figure 1b shows. More than 93% of all restarts come from rebuilds. Existing benchmarks miss the bursts because a percentile over a large batch cannot see a few thousand operations with high latency among millions. The known fixes trade one cost for another. The typical B-tree fix of randomizing each node's initial density spreads the rebuilds out but adds restarts under concurrency. At scale, smaller nodes lengthen runs and raise median lookup latency.

Two standard systems techniques, applied only where they are needed, remove most of the cost. Since a rebuild only reads the old node, inserts that reach it go to a small buffer that the rebuild merges into the new node, while lookups read the old node and the buffer. In addition, an insert that brings a node close to the threshold hands the rebuild to a background thread instead of performing it. Neither technique is new. Online index builds have long logged concurrent updates in a side file [25], while LSM trees and XIndex restructure in background threads [26, 29]. Unlike buffered learned indexes, which keep new keys outside the learned structure until they are merged, our buffer exists only while a node is rebuilt, so the index otherwise keeps its lookup path and memory.

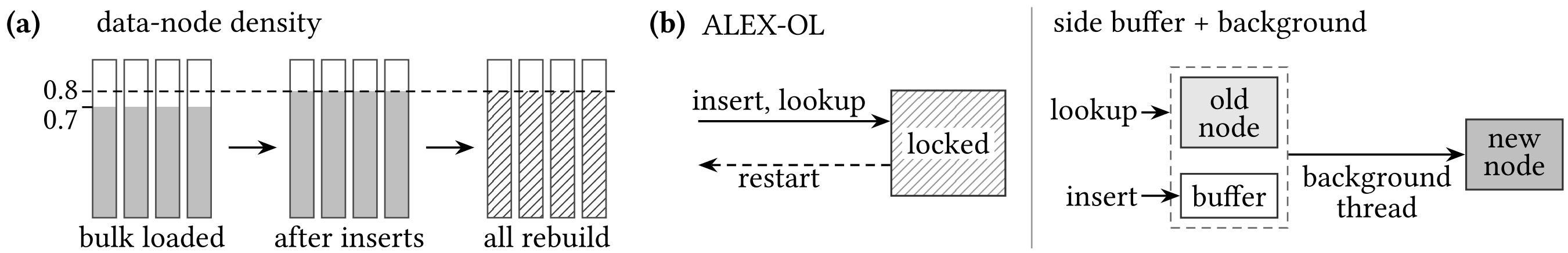

**Figure 1.** (a) Bulk loading fills every data node to the same density, so the nodes reach the expansion threshold together and rebuild in a burst. (b) In ALEX-OL a rebuild locks its node, and every insert and lookup that reaches it restarts until the rebuild ends. With the side buffer, inserts go to the buffer and lookups read the old node and the buffer, while a background thread builds the new node from both.

We apply both techniques to ALEX-OL, the concurrent gapped-array index of the GRE benchmark, which calls it ALEX+ [33], and measure it on 400M keys from the SOSD benchmark [10]. There, the bursts carry 89% of all inserts whose latency exceeds 100 µs, which we call *high-latency inserts*. The two techniques remove nearly all of them with one thread, and most of them with sixteen when spare cores run the background threads, while median lookup latency rises by 6–10%. On a second dataset, where the rebuilds are already spread out, the reduction is smaller. A permanent buffer in ALEX-OL itself makes the run about 17 times longer at sixteen threads.

The lesson is that the insert tail of a concurrent learned index is largely a systems problem. A recent robustness study attributes the instability of these indexes to overfitted models, unbalanced structures, ineffective adjustments, and excessive space reservation [21]. We locate the cost of concurrent inserts in synchronized restructuring and blocking, which existing benchmarks cannot see. The proper remedies are ones that concurrent data structures have long used.

We make the following contributions:

- We present a systems account of updatable learned indexes under concurrent inserts, on one common workload that checks every key: where each design pays for inserts, how that changes as the index grows, and why the standard benchmarks do not see it.
- We diagnose the most instructive case, in which bulk loading synchronizes rebuilds at positions that three source constants predict. We confirm the prediction by changing the constants, show that batch percentiles hide the bursts, and trace their cost to operations restarting on a locked node.
- We show that two standard techniques, applied only during a rebuild, remove most of this cost, which confirms the diagnosis. From this we draw lessons for index designers and benchmark authors.
- We release the artifacts behind every number: a benchmark driver that checks every key, patches that fix defects in released ALEX-OL, the run scripts, and the results of all runs, at https://github.com/mpi-dsg/GRE_Longitudinal/tree/wave-benchmark/wave-benchmark.

***Paper roadmap.*** Section 2 describes how learned indexes absorb inserts and what is known about waves of misery. Section 3 derives the burst positions, Section 4 shows why percentiles over large batches cannot measure them, and Section 5 locates their cost and applies the two techniques. Section 6 evaluates the designs, the bursts, the mitigations, and the techniques. Section 7 draws lessons and states limitations. Section 8 discusses related work.

## 2 Learned Indexes Under Inserts

***Learned indexes.*** A learned index answers the queries a B-tree answers over a sorted set of keys, but it replaces the comparisons at each level with a model, typically a linear function, that maps a key to its approximate position [11]. A lookup evaluates the model and then typically searches only a few slots around the prediction. The ideal case is a read-mostly workload over keys whose distribution a model fits well, such as the numeric identifiers and timestamps in the SOSD benchmark datasets [10]. There, a learned index can match a B-tree's lookup latency in much less memory. Inserts are the difficulty, because each insert shifts the positions a model was originally fitted to.

***Absorbing inserts.*** Updatable learned indexes absorb inserts in one of three ways. In a gapped-array index such as ALEX [5], keys are stored in *data nodes*, arrays kept partly empty, each with its own model, while a tree of *model nodes* routes each key to its data node. The *density* of a data node is the fraction of its slots in use. An insert writes into a free slot near the predicted position. Once a node's density crosses a threshold, the index replaces it with a larger node or with two smaller ones and refits the model. We call either replacement a *rebuild*. Model-placement indexes such as LIPP and SALI [7, 35] put each key at its predicted slot, create a child node when two keys collide, and rebuild a subtree once it has grown by a fixed factor. Buffered indexes such as XIndex and FINEdex [17, 29] keep new keys outside the learned array, in delta indexes or per-record bins that are later merged or retrained. All three are built by *bulk loading*, which fits the models over an initial sorted set of keys. Benchmarks construct them this way and then measure while inserting the remaining keys [28, 33].

***Synchronized restructuring.*** Bulk loading is known to synchronize restructuring in B-trees. A bulk-loaded B-tree fills every leaf page to the same fraction, so under uniformly distributed inserts the pages fill at the same rate and split at about the same time. The bursts of splits then typically recur, weaker each time, as the tree grows. Glombiewski et al. named these bursts waves of misery and proposed filling pages to different levels at bulk load, including a random fill, to spread them out [8]. Xing et al. measured them in R-trees and proposed remedies that act during inserts [36]. These studies measure costs such as leaf splits, buffer evictions, and I/O latency, but not the latency of concurrent operations.

***Why the waves are worse in learned indexes.*** A gapped-array index bulk loads every data node to one density, so it shows the same waves, which cost more for two reasons. First, the unit of restructuring is about 1000 times larger: a 512-byte $B^+$-tree page holds 31 entries, whereas a data node in the standard benchmark configuration holds up to 32,768, all of which a rebuild copies before refitting the model. Second, these indexes run with concurrent threads, so inserts and lookups that reach a node while it is rebuilt must wait or retry. The following sections establish where the bursts fall (Section 3), how to measure their cost (Section 4), and how to remove it (Section 5).

# 3 Why Bulk-Loaded Indexes Rebuild in Bursts

Section 2 argued that a bulk-loaded gapped-array index should rebuild its nodes in bursts. This section predicts where the bursts fall and states when they occur.

## 3.1 Predicting the Burst Positions

ALEX sets data-node density with three compile-time constants [34]. A bulk load fills a node to `kInitDensity_` ($d_{\text{init}} = 0.7$), a node is rebuilt into a larger one, an *expansion*, once its density reaches `kMaxDensity_` ($d_{\text{max}} = 0.8$), and expansion lowers the density to `kMinDensity_` ($d_{\text{min}} = 0.6$). Consider a data node bulk loaded with $n_0$ keys at density $d_{\text{init}}$, giving it capacity $n_0/d_{\text{init}}$. It expands once its key count reaches $d_{\text{max}} \cdot n_0/d_{\text{init}}$, that is, after its key count has grown by a factor of $d_{\text{max}}/d_{\text{init}}$. With the released constants the first expansion follows a growth of $0.8/0.7 = 8/7$. Expansion resets the node to $d_{\text{min}}$, so every subsequent expansion follows a growth of $d_{\text{max}}/d_{\text{min}} = 4/3$.

If the workload distributes inserts across nodes in proportion to the keys they already hold, every bulk-loaded node should follow this schedule at roughly the same time. Bursts therefore fall at key counts $N_0 \cdot 8/7$ and then at successive multiples of $4/3$, where $N_0$ is the number of keys bulk loaded, so the schedule can be computed in advance. For $N_0 = 10^8$ this predicts bursts at 114.3, 152.4, 203.2, 270.9, and 361.2 million keys.

Two properties let us test this account against measurements. First, the positions are fixed before measurement, since they follow from source constants and the bulk-load size alone. Second, the schedule is multiplicative, so the bursts spread apart as the index grows, which distinguishes it from any mechanism triggered by a fixed number of operations.

We confirm the premise directly. Immediately after bulk loading $10^6$ keys, every data node in the index sits at a density between 0.7000 and 0.7039 across three seeds.

## 3.2 Two Conditions for a Burst

Bursts require both a deterministic trigger and a set of nodes created at the same moment in the same state. LIPP-OL rebuilds a node once its size reaches twice its build size, which is as deterministic as ALEX's density threshold, and exhibits no bursts at any scale we measured. Its nodes are created at different times as collisions arise, which spreads their rebuild schedules.

The two conditions separate the three learned indexes. Only ALEX-OL satisfies both and exhibits bursts. SALI breaks at least the first by drawing its trigger randomly. Like LIPP-OL, it also creates nodes as keys collide. LIPP-OL breaks the second by creating nodes incrementally. Both controls are properties of released code.

This account attributes the bursts to uniform bulk-load density. Building ALEX without a bulk load would change more than the uniformity of its nodes, so the available test keeps the bulk load and removes only the uniformity, which Section 6.4 reports.

## 3.3 Index Size Reveals the Bursts

The burst schedule has one consequence that requires no latency instrumentation. Nodes that expand together make index size grow in discrete jumps, whereas unsynchronized nodes make it grow steadily. Over a 400M-key workload, ALEX-OL's median change in index size between consecutive batches is zero: sixteen of its 23 changes are below 0.05 GB, while six exceed 0.5 GB, up to 1.76 GB (the last burst spans two batches). LIPP-OL instead allocates between 1.30 and 1.40 GB in every batch.

## 4 Why Percentiles Miss the Bursts

Knowing where the bursts fall, we need a metric that can see them, because tail latency decides how an index behaves inside a larger service [4]. The cost of a burst is the high-latency operations it causes and the run time it adds. A burst produces roughly one high-latency insert per rebuilt node. A percentile reports the value at a fixed rank: the 99.9th-percentile latency (p99.9) of a million samples is the thousandth largest. A burst is therefore visible in p99.9 only if its high-latency inserts outnumber that rank. Uniform sampling scales both counts alike, so the batch size decides.

***Sampling leaves too few burst operations.*** The benchmark configuration we adopt times 1% of operations and reports percentiles over batches of $5 \times 10^7$ operations. Under a balanced read-write mix, that leaves about $2.5 \times 10^5$ timed inserts per batch, so p99.9 is the 250th largest of them. The first burst after a 50M-key bulk load rebuilds on the order of 4000 data nodes, half the 7781 we count after a 100M-key bulk load. Sampling one operation in a hundred records about 40 of the resulting high-latency inserts. Forty samples cannot reach a 250th-rank cutoff, so the burst barely moves that percentile.

***Large batches hide the bursts.*** Without sampling, the same failure occurs whenever the batch is large. On 400M-key runs, measured two ways with medians over three seeds: the p99.9 of batches of $1.25 \times 10^7$ operations that contain a predicted burst is 1.8× that of the other batches, which we call quiet batches. Counting inserts slower than 100 µs on the same data gives a ratio of 828×, which we call the *concentration* of the bursts.

Table 1 shows the effect of batch size on an independent set of runs, described in Section 6.1, in which the reporting batch is swept while everything else is held fixed. As the batch grows from 25M to 100M operations, the worst write grows 8.7-fold and p99.9 2.4-fold.

***Spreading the bursts looks like a cure.*** Where a burst's high-latency operations sit above the rank cutoff and the same operations spread thinly sit below it, a change that only redistributes work appears to remove it. Section 6.4 reports a 237-fold improvement of exactly this kind.

***What we count instead.*** We therefore time every operation and report how many exceed a latency of 10 µs, 100 µs, and 1 ms, alongside run time. The count of *high-latency inserts*, those above 100 µs, is our main tail-latency metric. Since every operation is timed, a count over a run does not depend on how the run is divided into batches, whereas counts from sampled latencies would scale with the sampling rate. Runs of different length must be compared per operation. Where we report a percentile, we state the batch size.

| ops/batch | write p99.9 (µs) | worst write (µs) |
|---|---|---|
| 25M | 6.8 | 310 |
| 50M | 13.4 | 750 |
| 100M | 16.0 | 2712 |

**Table 1.** ALEX on `libio` under the unmodified harness at three reporting batch sizes, three seeds. Same index, same workload, same machine: only the batch over which percentiles are computed differs. The harness samples one operation in a hundred. "Worst write" is the median over batches of each batch's largest sampled write latency.

## 5 Taking Rebuilds off the Critical Path

Sections 3 and 4 establish where the bursts fall and how to measure them, but not which operations pay for them. The *critical path* is the work an operation must wait for before it returns. To find out which operations pay, we instrumented ALEX-OL's insert path, which determines the design of our modification.

### 5.1 Who Waits for a Rebuild

ALEX-OL, the concurrent version of ALEX in the benchmark suite, follows the concurrency design of APEX [19, 33]. It protects model nodes with reader-writer locks and reads data nodes optimistically [15], validating each read against a version word. An insert takes its data node's lock with a single compare-and-swap. If the lock is held, the insert *restarts*. It abandons its traversal and starts again from the root. A rebuild, whether an expansion, a retrain, or a split, takes the same lock and holds it until the replacement node is published. Lookups validate against the same lock word and also restart while it is held. We tag each restart by what held the lock.

At sixteen threads, 97.9% of restarts on the 4M workload and 93.3% on 400M keys meet a node held by a rebuild. The rest meet an ordinary insert. Rebuilds holding node locks thus cause the node-level contention that Luo et al. report for this index [21]. At 400M keys a rebuild holds the lock for 440 µs on average and up to 8.0 ms. Every insert or lookup that reaches the node in that interval repeats its traversal until the interval ends. During a burst, most of the bulk-loaded nodes pass through that state within one batch.

Two properties of the rebuild make this blocking largely unnecessary. First, it only reads the old node's contents: it copies the keys into a new array and refits the model, while the old array does not change until it is reclaimed. Lookups can therefore read it safely for the whole interval. Second, the insert that crosses the threshold performs the entire rebuild before it returns. Detecting that a node needs restructuring and performing the restructuring are coupled on the insert path, although nothing requires them to be.

### 5.2 Design Overview

Our modification has one component for each of these two properties. Figure 1b shows both on a node that crosses the threshold during a burst. The *side buffer* reduces restarts by letting operations proceed against the old node and a small buffer while the rebuild runs, which is safe because the rebuild only reads the old node. *Background rebuilding* removes the rebuild from the insert path by letting an insert that brings the node close to the threshold hand the rebuild to a background thread and return. Neither component changes node size, layout, or the model, so lookups outside a rebuild follow the same path as before.

***Side buffer.*** Before a rebuild copies anything, the thread performing it publishes a small buffer of 1024 entries on the old node. An insert that finds the node locked and a buffer present appends its key to the buffer under a short spinlock and returns, after checking the old node and the buffer for a duplicate. A lookup that finds the node locked and a buffer present searches the old node and then the buffer. When the copy is complete, the rebuilding thread seals the buffer, inserts its entries into the replacement, and publishes the replacement as before. For a split, it routes each entry to the new node whose key range contains it, where a key on the boundary is also found by the neighbor check of Section 7. Inserts that arrive after sealing, or that find the buffer full, restart as they do today, so the window in which they wait covers the drain and the parent update rather than the copy. Searches of the old node use a variant of ALEX's search that updates no statistics, so they write nothing to it. The buffer is freed with the old node through ALEX-OL's existing epoch-based reclamation [9]. Lookups run inside ALEX-OL's epoch guard, so a lookup that reads the old node or its buffer finishes before either is freed.

The argument for correctness is short. While the buffer is open, only the rebuilding thread holds the node's lock. That thread never writes the old node's keys, payloads, occupancy bitmap, or model, which are everything a lookup reads. Appends and sealing both occur under the buffer's spinlock, so every append that succeeds precedes the seal, and every sealed entry is in the replacement before the replacement is reachable. A key whose insert has returned is therefore in the old node or the buffer, both visible to any lookup still holding the old node, or in the replacement once it is published. A buffered insert takes effect when its append completes under the buffer's lock, whereas a lookup served during a rebuild takes effect when it reads the buffer. Range scans do not read the buffer: as in released ALEX-OL, a scan that reaches a locked node restarts and so sees the replacement once it is published. Updates and deletes are unchanged. The argument also requires that nothing else writes the old node. In released ALEX-OL, *root expansion* did, when it widened the top of the tree for inserts that fall outside the key range the index was built for. We fix it (Section 7).

***Background rebuilding.*** An insert that takes a node past a soft threshold, density 0.75 against the hard threshold of 0.8, enqueues one of the node's keys and returns. One or more background threads take keys from the queue, locate the node that now holds each one, and run the same rebuild code the insert path would have run. If the node has already been rebuilt, it is skipped. If the queue falls behind and a node reaches the hard threshold first, the node is rebuilt by the insert path as before, so the change can never delay a rebuild.

Detection remains fully synchronized, since in a burst every node crosses the soft threshold together and is enqueued together. The queue, however, serializes their execution on the background threads, off the insert path.

Each component is insufficient alone, as Section 6.5 shows. The background thread without the side buffer moves the rebuild off the insert path, but the node stays locked while it runs, so inserts now wait for another thread instead of performing the work themselves. The side buffer alone lets other inserts proceed during copying, while the insert that crossed the threshold still performs the rebuild.

### 5.3 Implementation

Our modification is a C++17 patch of about 530 added lines to the ALEX-OL of the GRE benchmark, which also carries the defect fixes of Section 7 and the restart counters of Section 5.1. Both components are off by default and selected at run time, so one binary runs every configuration. The side buffer is a fixed array of 1024 entries, 16 KB, guarded by a spinlock. Background rebuilding uses a mutex-protected queue of keys served by one or more `std::thread` workers, which pause whenever the benchmark's batch timer is stopped.

## 6 Evaluation

We now test the account of Sections 3–5 with six questions, from the broad account to our modification.

**RQ1** Where does each kind of updatable learned index pay for inserts, and what do the standard benchmarks miss? (§6.2)

**RQ2** Does a bulk-loaded ALEX-OL restructure in bursts at the positions three source constants predict, and is synchronized expansion the cause? (§6.3)

**RQ3** Do the two available mitigations, randomized density and smaller data nodes, remove the cost, and if not, why? (§6.4)

**RQ4** Does taking restructuring off the critical path remove the cost? (§6.5)

**RQ5** How does the result compare with learned indexes designed for concurrent restructuring? (§6.6)

**RQ6** What does our modification cost in lookup latency, cores, and tuning? (§6.7)

### 6.1 Methodology

***Platform.*** We ran on four identical servers, each with two Intel Xeon Gold 6134M processors (16 cores, 32 hardware threads) and 755 GB of memory, running Debian with GCC 14.2 and oneTBB, with nothing else running. Every table, figure, and comparison built from our runs uses one server and one driver. Each uses its own paired set of runs, so a configuration can differ slightly between sections.

***Indexes.*** We drive every index through the index interface of the GRE benchmark [33], so all of them follow one code path. The learned indexes are ALEX-OL, LIPP-OL, SALI, XIndex, and FINEdex, where the suffix -OL marks a concurrent version with optimistic locking. The classical controls are ART-OLC [14, 15] and a B$^+$-tree with optimistic lock coupling (OLC) [15, 32] and 31 entries per page, which we bulk load with our own loader since GRE's wrapper builds it by ordinary inserts. The driver runs threads through OpenMP because LIPP-OL and SALI index per-thread state by OpenMP thread number, without which SALI races silently. XIndex and FINEdex need one least-squares routine from Intel's Math Kernel Library (MKL), which our servers lack, so we substituted an implementation that solves the same problem exactly up to rounding.

***Workloads and metrics.*** Runs at 400M keys bulk load $10^8$ SOSD `books` (book popularity) or `osm` (OpenStreetMap cell identifiers) keys [10] and insert $3 \times 10^8$ more in 24 batches, either insert-only or with half or 90% of the operations replaced by lookups of bulk-loaded keys. Runs at 4M keys bulk load $10^6$ keys of a shuffled dense sequence and insert $3 \times 10^6$ in batches of $10^5$. Unless a caption says otherwise, we report medians over three seeds at 400M and over nine runs (three seeds, three repetitions) at 4M. Time is the sum of batch times, with background threads paused while the batch timer is stopped. Memory is the index's own report, or peak resident set size where indexes are compared, since several report none. Every run reads back every inserted key and compares its payload, a check that no ALEX-OL run failed. The 400M restart counters, the 400M lookup latencies of RQ3, and the one-thread and sorted insert orders at 400M use one seed. The 4M attribution of restarts to rebuilds and soft thresholds other than 0.75 use three runs.

***Independently collected runs.*** We also analyze runs collected with the unmodified benchmark harness before this mechanism was identified: 200M-key tables on five datasets with varying bulk-load fraction $\rho$ and read-heavy, balanced, and write-heavy mixes, 800M-key configurations, mixes with updates and deletes, and the batch-size sweep of Table 1. Since they sample 1% of latencies, we use them for out-of-sample tests, for behavior as the table grows, and for questions our runs cannot answer.

### 6.2 RQ1: Where Each Design Pays

***As the table grows.*** The independently collected runs follow ALEX, LIPP, DILI, ART, and a B$^+$-tree from a bulk load of 50M keys to 200M under a fixed mix, with three seeds per dataset. Under a read-heavy mix on one thread, the learned indexes start well ahead of the classical ones: on `libio`, ALEX serves 3.51M operations per second against 1.63M for ART and 1.12M for a B$^+$-tree. By 200M keys, ALEX has lost 25% of that throughput, LIPP 50%, and DILI 42%, while ART has gained 3%. On `osm`, ART leads at all but one table size. Under a balanced mix with 32 threads, averaged over four datasets, ALEX-OL falls from 36.7M to 27.8M operations per second and SALI from 36.7M to 26.5M, while ART-OLC starts at 43.8M and ends at 44.6M. The learned indexes' advantage is therefore largest right after the bulk load, which is typically where benchmarks measure it.

***Where the cost goes.*** Table 2 runs every concurrent index we could build on one 400M-key workload. There is no design that comes out ahead on every column. ALEX-OL has the lowest run time and the lowest lookup latency under the read-heavy mix, but its high-latency inserts arrive in the bursts the rest of this section examines. The model-placement indexes spread theirs fairly evenly across every batch and use 3.5 and 4.2 times ALEX-OL's peak memory, with LIPP-OL also taking 6.0 times as long. The buffered indexes incur relatively few high-latency inserts at 2.0 and 2.7 times the run time and 1.7 and 2.2 times the memory. Among the classical indexes, the B$^+$-tree incurs few high-latency inserts and uses the least memory at 1.5 times ALEX-OL's run time, while ART-OLC incurs the most of any index, spread evenly rather than in bursts. What sets ALEX-OL apart is that its high-latency inserts are synchronized, which makes them predictable and removable.

***Correctness.*** Our driver reads back every inserted key and its payload after each run. The standard harness instead counts the success flags that each index's wrapper returns, with its check of the values that lookups return commented out [34]. Three of the five concurrent learned indexes failed our check or stopped making progress. Two of them lost keys without crashing, which no success flag reveals. Released ALEX-OL lost up to 92% of a sample of its keys when inserts extended its key range early (Section 7). FINEdex lost one or two keys in two of nine runs at 400M keys with sixteen threads, while XIndex made no progress within 40 minutes in four of nine such runs and in one of two on `osm` within 25 minutes. Both run with our substitute for MKL's least-squares routine, which we cannot rule out as a factor, although Luo et al. also report crashes and deadlocks in concurrent learned indexes [21]. The harness wrappers have defects of their own: the ALEX and STX B$^+$-tree wrappers fail every update without calling the index, while the ART

| Index | design | insert-only >100 µs | insert-only time (s) | insert-only RSS (GB) | lookup (ns) 50/50 | lookup (ns) 90/10 | key check |
|---|---|---|---|---|---|---|---|
| ALEX-OL | gapped array | 70,928 | 17.2 | 18.0 | 536 | 456 | passed |
| + side buffer, background | gapped array | 4,029 | 17.3 | 18.6 | 567 | 456 | passed |
| LIPP-OL | model placement | 59,047 | 102.6 | 63.4 | 1668 | 906 | passed |
| SALI | model placement | 83,879 | 38.0 | 75.8 | 820 | 678 | passed |
| XIndex | buffered | 4,793 | 45.7 | 30.3 | – | 763 | hung 4/9 |
| FINEdex | buffered | 5,753 | 33.8 | 40.3 | 894 | 873 | lost keys 2/9 |
| $B^+$-tree-OLC | classical | 6,943 | 25.5 | 15.4 | 1066 | 918 | passed |
| ART-OLC | classical | 103,187 | 21.8 | 29.4 | 722 | 718 | passed |

**Table 2.** Every concurrent index on one workload: 400M SOSD books keys, bulk 100M, then 300M operations at sixteen threads. ">100 µs" counts high-latency inserts, time is the insert-only run time in seconds, RSS is peak resident memory in GB (about 5 GB of it is the driver's key arrays, common to all), and lookup is the median latency under a balanced and a read-heavy mix. The key check reads back every inserted key. ALEX-OL includes our fixes to its released defects (Section 7). Values are medians of three seeds, or of the completed runs for XIndex.

wrapper reports every delete as successful whether or not it removed a key.

In summary, each design pays for inserts somewhere. Gapped-array indexes pay in synchronized bursts, model-placement indexes in memory, buffered indexes in run time, and several implementations in correctness. The rest of this section examines the bursts.

### 6.3 RQ2: Predicted Bursts and Their Cause

RQ2 tests the premise on which the rest of the diagnosis depends, namely that the bursts fall where Section 3 places them and for the reason it gives.

***At scale, on real keys.*** Figure 2 shows ALEX-OL bulk loaded with $10^8$ SOSD books keys and given $3\times10^8$ inserts in 24 batches on one thread, with the predicted positions shaded. The predicted batches carry a median of 8284 high-latency inserts, against ten in the remaining batches. Across the run, 89% of all high-latency inserts fall in the five predicted batches, which are 21% of the batches. Index size steps at the same five points, from 2.31 to 9.72 GB. This concentration of 828× becomes 677× at sixteen threads, so concurrency does not sharpen the bursts. LIPP-OL, SALI, the $B^+$-tree, and ART-OLC show no concentration, with ratios of 0.8 to 1.0, and place 16–21% of their high-latency inserts in the predicted batches at one thread, roughly what an even spread would produce.

***What a burst costs a workload.*** A burst also slows every other operation while it lasts. At 400M keys, a batch that contains a predicted burst takes 13.5 s at one thread against 8.0 s for a quiet batch, and 0.98 s against 0.64 s at sixteen threads, so throughput falls by 35–41% over those batches. At sixteen threads, such a batch holds a median of 271 inserts slower than 1 ms, against none in a quiet batch. Its slowest insert takes 7.7 ms against 0.68 ms. A service built on the index would see stalls of several milliseconds that recur at predictable points as the index grows.

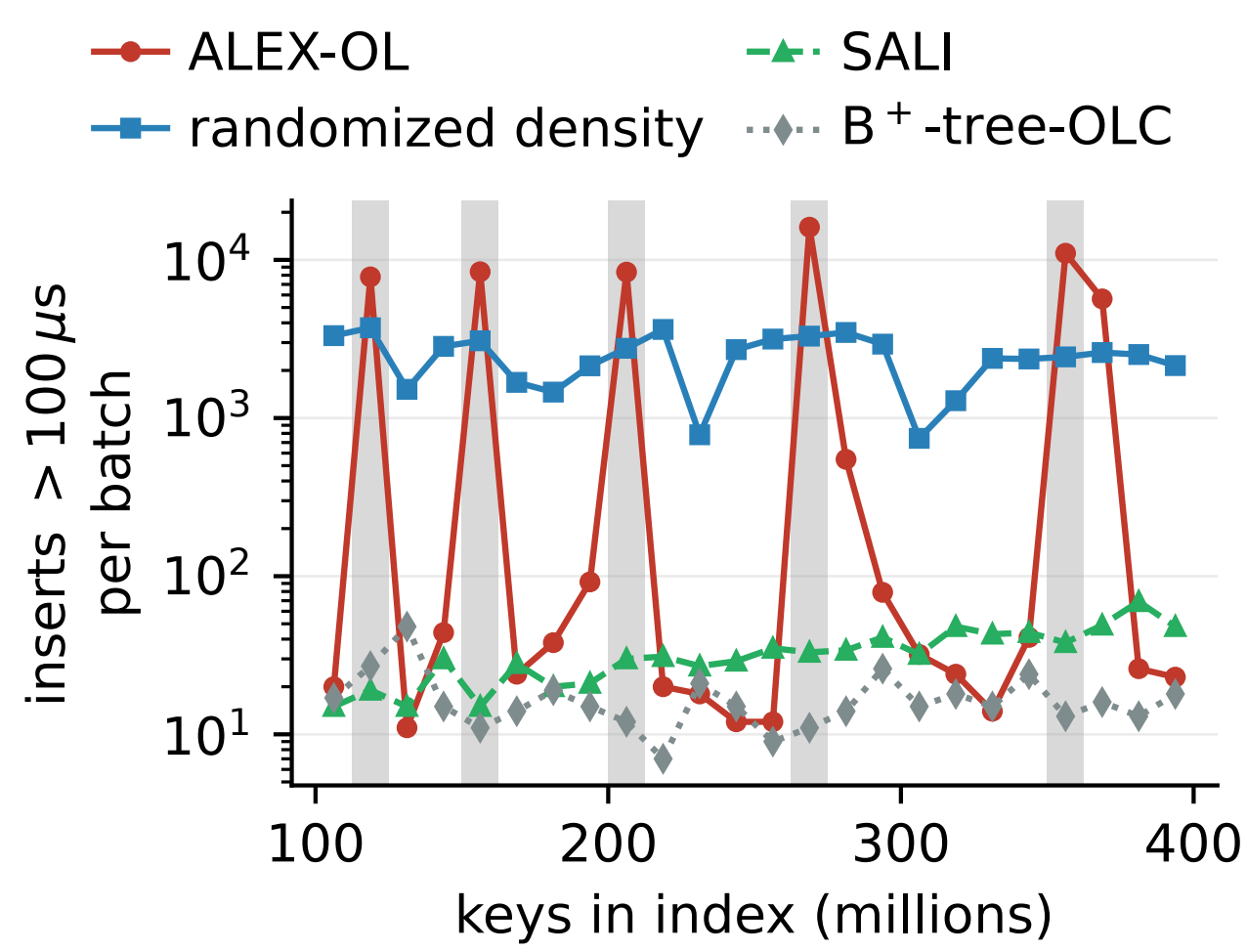

**Figure 2.** Inserts exceeding 100 µs per batch on 400M SOSD books keys, one thread, seed 1866. Shaded bands mark the batches that contain a position predicted from three source constants. "Randomized density" is ALEX-OL with a randomized bulk-load density.

***Across thread counts and mixes.*** At 4M keys, from one to sixteen threads, ALEX-OL's ratio of worst-batch to median-batch p99.9 stays between 269× and 562×, while LIPP-OL, SALI, and the $B^+$-tree stay between 1.0× and 10.7×. ART-OLC is the one index whose ratio grows steadily with concurrency, from 2.2× to 61.4×, with its high-latency inserts spread across every batch. In the independently collected balanced and read-heavy mixes on `libio`, `planet`, and `osm`,

ALEX's write p99.9 at predicted batches is 2.2 and 4.4 times that elsewhere.

***Changing the constants moves the bursts.*** The most direct test of the cause changes the constants themselves. We rebuilt ALEX-OL with $d_{\text{init}} = 0.6$, $d_{\text{max}} = 0.8$, and $d_{\text{min}} = 0.5$, which moves the first burst to $N_0 \cdot 4/3$ and spaces the later ones by a factor of 1.6, and recorded the predicted batches before running it: 4, 12, and 25 at 4M keys, and 3, 10, and 20 at 400M. In all eight runs, at one and sixteen threads, the three largest bursts fall in exactly those batches. With the released constants, the same configurations burst in batches 2, 6, 11, 18, and 27 at 4M keys and one thread, and 2, 5, 9, 14, and 21 at 400M, again as predicted.

***Out-of-sample: moving the bulk-load size.*** The independently collected runs, gathered before the mechanism was identified, vary the bulk-load fraction $\rho$ on a 200M-key table, where each $\rho$ predicts a different schedule: the first burst falls at 22.9M keys for $\rho = 0.10$, at 57.1M for 0.25, and at 114.3M for 0.50. Pooled over two datasets and three seeds, batches that contain a predicted burst show a write p99.9 of 18.7 µs against 9.0 µs elsewhere ($n = 102$ batches) and an index-size step of 0.63 GB against 0.20 GB. The contrast is 1.2× at $\rho = 0.10$, where 36 of 42 batches contain a burst, and 2.6× and 2.5× at $\rho = 0.25$ and 0.50.

***Where the bursts weaken.*** Bursts require nodes to fill at similar rates, which requires a reasonably even key density, so they should weaken where density is uneven. On `osm` at 400M keys, the concentration falls from 828× to 2.6×, while the share of high-latency inserts in predicted batches falls from 89% to 40%, against 21% for an even spread. The quiet batches carry a median of 647 high-latency inserts, against ten on `books`, so the restructuring cost is spread across every batch while the total, 35,225, stays of the same order as the 57,450 on `books`. This suggests that an uneven key distribution redistributes restructuring much as randomized density does. An index without bursts can still be expensive. SALI has none on either dataset, yet incurs 498,233 high-latency inserts on `osm`, fourteen times ALEX-OL's count.

***A B⁺-tree with randomized fill.*** For a bulk-loaded B⁺-tree with leaves filled uniformly to 0.70 or 0.90, or randomized around either, the ratio of worst to median batch p99.9 stays between 1.2× and 2.1× from one to sixteen threads. At a uniform fill of 0.90 its leaves split within a few inserts of one another, which is the configuration waves of misery describe, yet each wave stays cheap. The two indexes thus share the mechanism but differ in the cost per event. The B⁺-tree also shows that the cost randomized density adds under concurrency (Section 6.4) is not a general property of optimistic validation.

In summary, ALEX-OL places 89% of its high-latency inserts in the five batches that three source constants predict. The bursts move with the constants or bulk-load size, while the other indexes show no comparable latency bursts.

| | | inserts > 100 $\mu$s | | | p99.9 |
|---|---|---|---|---|---|
| Thr | density | total | wave | quiet | worst |
| 1 | released | 594 | 90 | 0 | 181.6 |
| 1 | randomized | 597 | 20 | 19 | 0.77 |
| 16 | released | 4407 | 687 | 0 | 1068.3 |
| 16 | randomized | 7061 | 271 | 233 | 1048.0 |

**Table 3.** Randomized bulk-load density against baseline, $10^6 \rightarrow 4 \times 10^6$ keys, nine runs per cell. Counts are inserts slower than 100 µs. The p99.9 column reports what a percentile says about the same runs.

### 6.4 RQ3: Do the Available Mitigations Remove the Cost?

We test the mitigation established for B-trees, randomizing the bulk-load density, and the one parameter ALEX exposes that bounds the cost of a rebuild, the data-node size.

***Randomized density spreads rebuilds out.*** Randomizing the bulk-load density can be implemented with a single line in ALEX. Instead of setting every node to the initial density, we draw a capacity multiplier uniformly from $[1/d_{\text{init}} - \delta,\ 1/d_{\text{init}} + \delta]$ and invert it, which keeps the expected capacity equal to the released configuration. This spreads the rebuilds out as intended. Node densities span 0.626–0.796 after the change against 0.7000–0.7039 before it. The share of high-latency inserts in predicted batches drops from 80% to 20% at 4M keys, close to the 17% an even spread would produce, and from 89% to 26% at 400M, against 21%.

The total cost, however, stays the same. As Table 3 shows, at one thread the randomized configuration incurs 597 high-latency inserts against the baseline's 594 at 4M keys, and 58,989 against 57,539 at 400M. The same restructuring still occurs, only redistributed: the quiet-batch count rises from zero to 19 at 4M keys and from nine to 2398 at 400M.

***A percentile reports a false improvement.*** Measured by p99.9 over batches of $10^5$ inserts, the same runs report worst-batch latency falling from 181.6 µs to 0.77 µs, a 237-fold improvement, although no cost was removed. The baseline's median burst batch holds 90 high-latency inserts, while its worst holds at least the 100 that p99.9 needs to report a high value. The randomized configuration spreads the same operations at about 20 per batch, below that cutoff everywhere, so the percentile reports a value from the low-latency body of the distribution. A metric too coarse to observe the bursts can thus report an improvement that leaves them in place.

***Under concurrency randomized density adds cost.*** At sixteen threads, the randomized configuration incurs 7061

high-latency inserts against the baseline's 4407, an increase of 60%, and at 400M keys 85,741 against 71,517, an increase of 20%. We ruled out NUMA placement, since pinning threads leaves the effect intact, and nearly full gapped arrays, since a one-sided draw behaves like the symmetric one. A narrower draw gives an intermediate result, so the penalty scales with the spread.

***Restructuring forces inserts to restart.*** We counted the restarts of ALEX-OL's insert path. At sixteen threads, the baseline restarts about $1.5 \times 10^5$ times per batch of $10^5$ inserts, but $9.8 \times 10^6$ times at the first predicted burst, about 98 restarts per insert issued. At one thread the counter reads zero in both configurations, which is why randomized density is free at one thread and costly at sixteen. Over a run, randomized density raises total restarts from $1.37 \times 10^8$ to $2.57 \times 10^8$, a factor of 1.87, while high-latency inserts rise by a factor of 1.6. Section 5.1 attributes nearly all restarts to nodes locked by a rebuild. Our reading, consistent with these counters but not a measured law, is that under a retry-based scheme the delay of an arriving insert depends less on the peak rate of rebuilds than on how likely it is to meet one, and spreading rebuilds out raises that likelihood. This trade favors a single-threaded B-tree, where a split costs only the thread performing it, but not an index in which a rebuild delays every thread that arrives during it.

***Smaller data nodes.*** Smaller data nodes should shorten each rebuild. We swept the bound from 32 to 131,072 entries per node at 4M keys and repeated the most effective setting at 400M. At 4M keys, 2048 entries instead of the standard 32,768 take high-latency inserts from 605 to 10 at one thread and from 4471 to 933 at sixteen, as Figure 3 shows, while the run takes 31% less time at sixteen threads. The cost falls on the read path, which is what large data nodes buy: under a balanced mix, median lookup latency rises from 102 to 150 ns at one thread. As Table 4 shows, the gain shrinks at 400M keys. High-latency inserts fall from 57,450 to 6849 at one thread but only from 71,367 to 51,554 at sixteen. At one and sixteen threads, the run takes 37% and 34% longer. Median lookup latency rises by 61% and 34%, since with sixteen times as many nodes the tree is deeper. At sixteen threads, most high-latency inserts remain, for reasons we have not isolated.

In summary, neither mitigation removes the cost at scale. Randomized density redistributes it and adds 20–60% under concurrency, while smaller nodes leave rebuilds on the critical path and make the run take 34–37% longer.

### 6.5 RQ4: Removing the Bursts from ALEX-OL

If the cost of a burst lies in restructuring on the critical path (Section 5), taking it off the critical path should remove most of the cost while as much restructuring occurs. We evaluate the side buffer and background rebuilding separately and together, with one, two, or four background threads.

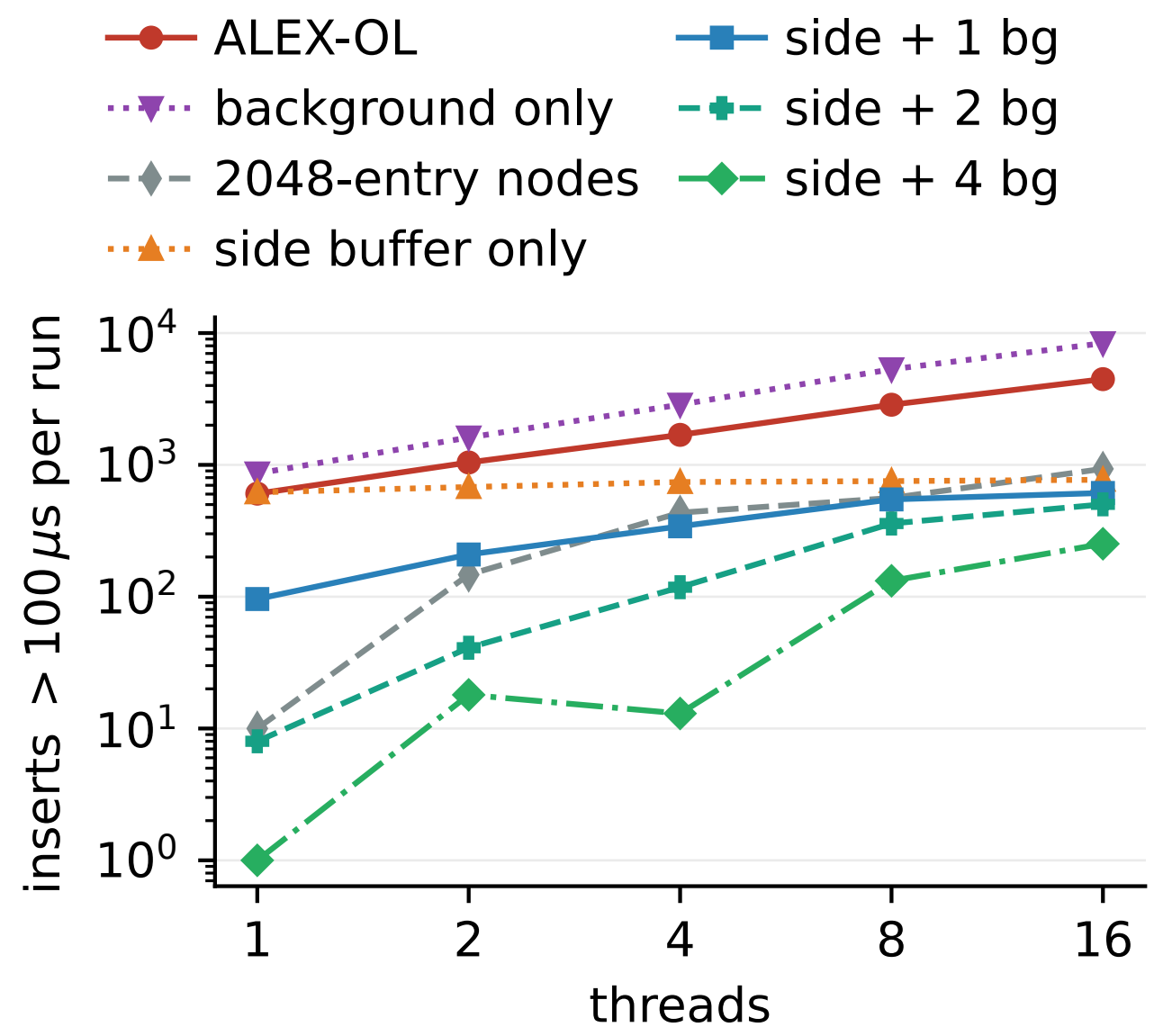

**Figure 3.** Inserts slower than 100 µs per run at 4M keys, median of nine runs per point, all configurations from one binary. "Background only" rebuilds on a background thread without the side buffer, and "side + $k$ bg" is the side buffer with $k$ background threads.

***Across thread counts.*** Figure 3 reports the 4M workload. With the side buffer and one background thread, high-latency inserts fall from 605 to 96 at one thread and from 4471 to 612 at sixteen. With four background threads they fall to 1 and 252, a factor of 18 at sixteen threads, while the run takes 35% less time. The ablations separate the two changes as Section 5.2 predicts. The side buffer alone leaves the single-threaded count unchanged, since with one thread no other insert is waiting, but reduces it to 771 at sixteen threads. Background rebuilding alone is worse than the baseline at every thread count, with 8372 at sixteen threads, because the node stays locked while the background thread rebuilds it.

***At scale.*** Table 4 and Figure 4 report the 400M workload. At one thread, the side buffer with one background thread reduces high-latency inserts from 57,450 to 249 and the worst batch from 15,886 to 26, while the run takes 9% less time. At sixteen threads, one background thread reduces high-latency inserts only to 31,746, two to 12,407, and four to 4315, a factor of 17, at an unchanged run time of 17.2 against 17.3 s. The reduction is smaller in the extreme tail: inserts over 1 ms fall from 1227 to 276, while inserts over 10 µs fall by 25%. Batches that contain a predicted burst now take as long as quiet ones, 0.70 against 0.72 s. On `osm`, where the cost is already spread across every batch, the reduction is smaller, from 35,225 to 26,734 high-latency inserts at one thread and from 94,218 to 35,072 at sixteen with four background threads.

| Configuration | 1 thread | | | | 16 threads | | | |
|---|---|---|---|---|---|---|---|---|
| | >100 µs | worst | time | mem | >100 µs | worst | time | mem |
| ALEX-OL | 57,450 | 15,886 | 214 | 9.72 | 71,367 | 17,502 | 17.3 | 9.72 |
| 2048-entry nodes | 6,849 | 1,273 | 293 | 9.83 | 51,554 | 11,561 | 23.2 | 9.83 |
| side buffer only | 57,780 | 16,078 | 215 | 9.72 | 58,812 | 16,160 | 16.8 | 9.72 |
| side buffer + background | 249 | 26 | 194 | 8.97 | 31,746 | 7,469 | 18.0 | 9.64 |
| side buffer + 2 background | – | – | – | – | 12,407 | 5,454 | 18.3 | 9.69 |
| side buffer + 4 background | – | – | – | – | 4,315 | 3,272 | 17.2 | 9.56 |

**Table 4.** 400M SOSD books keys, bulk 100M then 300M inserts in 24 batches, median of three seeds, all configurations from one binary. ">100 µs" counts high-latency inserts across the run, "worst" is the largest count in any one batch, time is in seconds, and memory is the index's own report in GB at the end of the run. Two and four background threads are measured at sixteen threads only.

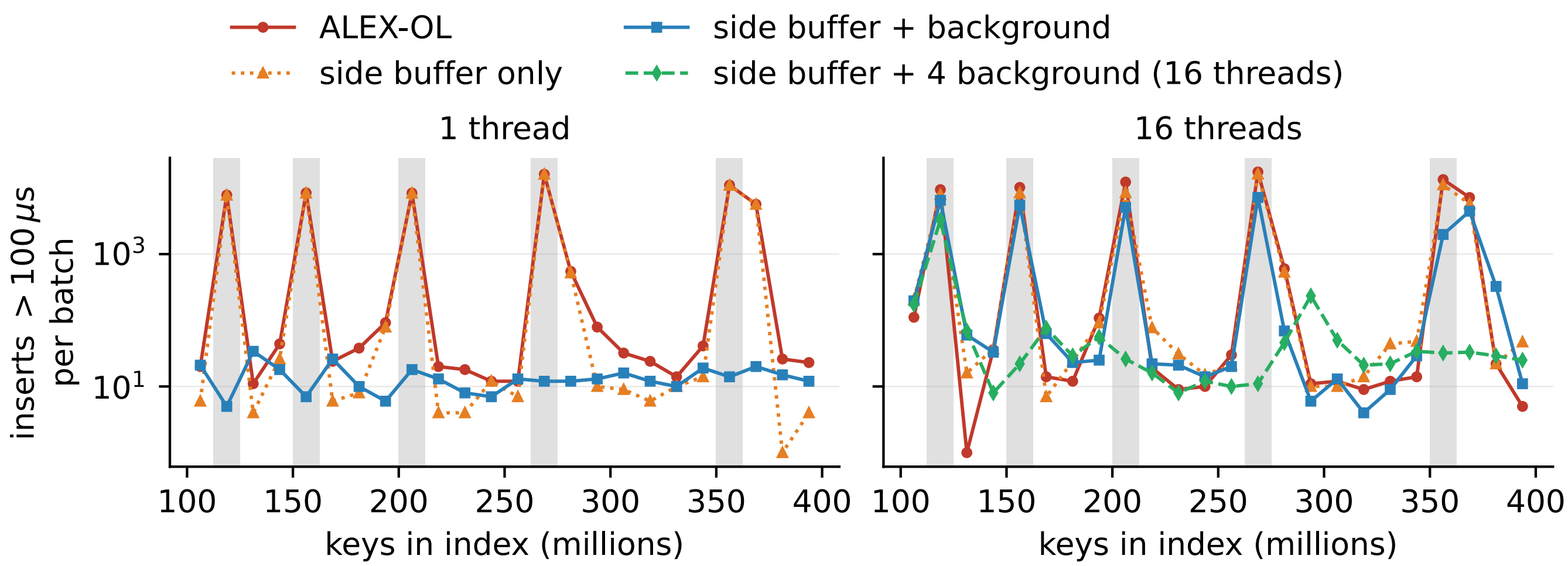

**Figure 4.** Inserts slower than 100 µs per batch on 400M books keys, seed 1866. Shaded bands mark the batches that contain the baseline's predicted bursts.

***Insert order.*** The workloads above insert keys in random order. When inserts concentrate on hot key ranges, drawn from 1024 ranges with Zipfian weights, the baseline incurs more high-latency inserts at 400M keys and sixteen threads, 112,725, of which modified ALEX-OL removes nearly all, leaving 2040 at the same run time. At one thread they fall from 59,208 to 1147. Sorted inserts, however, are a clear exception. Every insert then reaches the rightmost data node, which is rebuilt about every 130 inserts, so there are no bursts to remove. There, modified ALEX-OL does worse, raising high-latency inserts from 59,272 to 94,112 at sixteen threads while the run takes 40% longer. The counters show that the side buffer fills during each rebuild of that node, so inserts that find it full restart, 46 million times over the run, while the background threads add rebuilds to a node the foreground already rebuilds continuously.

***Where the remaining high-latency inserts come from.*** We classified every high-latency insert at 400M keys with sixteen threads. In the baseline, 57,030 performed a rebuild themselves, while 12,200 restarted on a node under rebuild. With the side buffer, restarts on a rebuild disappear. With one background thread, 94% of the remaining high-latency inserts performed a rebuild because no background thread had reached their node before the hard threshold. With four, 54% did. One background thread performs 53% of the rebuilds, whereas four perform 97%. The high-latency inserts track the remaining foreground rebuilds: 29,138 with one background thread and 2470 with four. Our modification, therefore, needs background capacity roughly in proportion to the foreground insert rate.

In summary, on books, taking restructuring off the critical path cuts high-latency inserts by 231× at one thread and by 17× at sixteen, given enough background threads to keep pace with the foreground inserts.

### 6.6 RQ5: Comparison with Indexes Built for Concurrent Restructuring

XIndex and FINEdex were designed to restructure while serving concurrent operations [17, 29]. If they already avoided bursts at no cost, our proposed modification would add little. Both avoid bursts but cost more elsewhere (Table 2). At sixteen threads, XIndex takes 2.7 times as long as ALEX-OL for the same inserts in its completed runs, FINEdex 2.0 times. Under a 90/10 mix, their median lookups take 1.7 and 1.9 times as long. XIndex completed one of six runs under a balanced mix, none of them in Table 2. Their peak memory is 1.7 and 2.2 times as large. ALEX-OL with the side buffer and four background threads incurs fewer high-latency inserts than either, 4029 against 4793 and 5753, at ALEX-OL's run time and with 3.5% more peak memory.

***A permanent buffer in the same index.*** To compare within ALEX-OL, we also measured a variant in which every data node always keeps a buffer of 1024 entries. Every insert goes to the buffer, a lookup searches the array before the buffer, and the inserting thread merges a full buffer into its node through ALEX's own rebuild path. On one server at 400M keys and sixteen threads, this variant takes 284 s for inserts that take 17.0 s in the baseline and 18.0 s with our modification, while incurring 335,334 high-latency inserts. Median lookup latency under a balanced mix rises from 538 to 642 ns, while memory rises by 7%. With four background threads it still takes 212 s. At one thread it takes 516 s against 213 s for the baseline and 189 s with our modification. At 4M keys with sixteen threads, 16 of its 18 runs did not finish within two minutes, whereas the baseline takes 0.23 s. Confining the buffer to a rebuild therefore appears to keep its cost small. This variant is only one way to buffer always, with merges on the insert path. Other buffered indexes may cost less.

Our modification reduces ALEX-OL's burst costs. Modified ALEX-OL beats XIndex and FINEdex on high-latency inserts at a third to half of their run time, at a small fraction of a permanent buffer's cost.

### 6.7 RQ6: What the Modification Costs

Our modification adds work to lookups that reach a rebuild, uses background threads, and has one parameter, the soft threshold.

***Lookups.*** Since node size stays the same, lookups change little. Under a balanced mix at 400M keys, median lookup latency is 567 against 534 ns at sixteen threads with four background threads, and 412 against 374 ns at one thread, whereas 2048-entry nodes raise it to 724 and 602 ns, respectively (one-thread and 2048-entry values from one seed). Lookups slower than 100 µs at sixteen threads fall from 5204 to 180, because a lookup that reaches a node under rebuild now reads the old node instead of restarting. High-latency inserts fall from 29,775 to 1223. These lookups target bulk-loaded keys. When they instead target keys the same thread has just inserted, which reach nodes under rebuild and their buffers, every lookup found its key, 23,812 of them through a side buffer in one run. Median lookup latency then rises by 8%, while lookups over 100 µs fall from 4211 to 193.

***Cores and CPU time.*** The background threads use cores that the baseline would otherwise leave idle. In one set of runs at 400M keys with sixteen threads, they add 6% to the CPU time (290 against 273 s) and leave the run time unchanged, while with one thread they add 9% CPU time. At an equal budget of sixteen threads, twelve inserting and four in the background, high-latency inserts fall from 71,408 to 3138, while the run takes 22% longer since only twelve threads insert. With 32 foreground threads, which occupy every hardware thread, and four background threads, high-latency inserts at 400M keys fall from 93,315 to 42,366 over three seeds, a factor of 2.2 rather than 17, with a 15% longer run.

***Soft threshold.*** The soft threshold matters most at one thread. At densities of 0.68, 0.72, 0.75, and 0.78, high-latency inserts at 4M keys are 2, 15, 96, and 239 at one thread, and between 612 and 661 at sixteen.

## 7 Lessons and Limitations

***Lessons for index designers.*** The costs we measured come from how an index restructures. Keeping restructuring off the critical path only while a node is rebuilt removed most of that cost at a small price in lookup latency and memory. Any index that bulk loads its nodes to one fill level and rebuilds them at a fixed threshold will synchronize, whether or not its nodes hold models. The cost then grows with the size of the unit it rebuilds and with how long a rebuild blocks other operations. Designers can spread the triggers, shrink the unit, or move the rebuild off the critical path. In our measurements, only the last removed most of the cost while keeping run time and lookup latency close to the baseline.

***Lessons for benchmark authors.*** Three practices would have exposed what we found. First, count operations above fixed latency thresholds or keep full histograms [31], since percentiles over large batches cannot see thousands of high-latency operations among millions (Section 4). Second, report how performance changes as the index grows, since the learned indexes' lead is largest right after the bulk load. Finally, read back every inserted key after a run, since only that reveals lost keys.

***Scope of the modification.*** Our modification helps most where rebuilds are synchronized. Under sorted inserts, it adds cost (Section 6.5). The side buffer covers inserts and point lookups, while scans, updates, and deletes still restart on a node under rebuild and are not in our workloads. The background threads need spare cores (Section 6.7). ALEX-OL is the only index we modified. Our account of the cost assumes that an operation meeting a rebuild restarts (Section 6.4). A blocking scheme may scale differently, although the side buffer should still remove the wait.

***Correctness.*** Our exhaustive key check exposed three defects in released ALEX-OL, which we fixed in every configuration, the baseline included. First, root expansion (Section 5.2) left each child's recorded depth unchanged, so the child's next rebuild overwrote its siblings' pointers and detached their subtrees. It also modified the live root in place and freed the old child array while lock-free lookups could still hold it, so we now build the expanded root privately and publish it with one store. Second, a rebuild copied its leaf's depth before locking the parent, so a concurrent root expansion left it stale. Third, on `osm` keys, a split placed a key on a floating-point boundary in the right child while lookups routed it to the left one, so lookups that miss now check the adjacent leaves. With a bulk load of $10^5$ keys at 32 threads, where root expansion happens early, released ALEX-OL crashed in one of eight runs and lost between 19% and 92% of a sample of keys in three others. With the fixes, all 120 runs of the baseline and modified ALEX-OL at 8, 16, and 32 threads found every key. No instrumented timed run expanded the root.

***Scope of measurement.*** Our measurements come from one machine model and one implementation of each index, with the model-node bound held at its default throughout. The independently collected runs used a different harness configuration, so they support the positional claims but not the absolute counts. Since our driver is closed-loop, a thread waiting on a rebuild issues no other operation, so an open-loop client would see more high-latency operations than we count, an effect known as coordinated omission [27, 30]. Finally, we explain the weakening on `osm` only qualitatively.

## 8 Related Work

Our work sits between studies of restructuring bursts in classical indexes, designs and benchmarks of updatable learned indexes, and techniques for restructuring without blocking.

***Waves of misery.*** Split bursts after bulk load were characterized for B-trees [8] and R-trees [36] (Section 2). We add latency under concurrency, where randomized density adds cost (Section 6.4), but did not test insert-time remedies.

***Updatable learned indexes.*** ALEX introduced gapped arrays with per-node models [5]. LIPP made lookups exact by placing keys at predicted positions [35]. SALI added probability-based adjustment for concurrent workloads [7]. DILI, FINEdex, XIndex, and PGM explore related points [6, 17, 18, 29]. When integrated into Bigtable, learned indexes are built for immutable SSTables as these are written, so the indexes never absorb inserts [1].

***Benchmarking learned indexes.*** SOSD standardized the datasets and established the read-only advantage of learned indexes [10, 23]. GRE supplies the harness and much of the configuration we adopt, and reports p99.9 latency over sampled operations [33]. It also finds that background threads can raise latency variance. Our background rebuilds help most when cores are spare (Section 6.7). GRE's harness counts the success flags its wrappers return without reading back what was inserted (Section 6.2). Later studies examine learned indexes under concurrency, on disk, and across workloads [12, 28, 37]. Luo et al. are closest. They time every operation, report failures of concurrent implementations, and observe that the insert tail of concurrent ALEX comes from contention on node latches [21]. We show where that contention comes from: bulk loading synchronizes rebuilds, each of which holds a node's lock while other operations restart. We predict where the rebuilds fall, show that batch percentiles hide them, find keys lost without a crash, and remove most of the cost within ALEX-OL.

***Restructuring without blocking.*** Concurrent trees have long let operations proceed while a node is restructured. B-link trees let a reader traverse a split in progress through a right link [13], the Bw-tree applies updates to a node as delta records that are consolidated later [16], Masstree's readers take no locks [22], and BzTree installs restructured nodes with a multi-word compare-and-swap [2]. Index builds have long logged concurrent updates in a side file and applied them before completing [25]. Among learned indexes, XIndex buffers writes in per-group delta indexes that background threads merge [29], FINEdex places new keys in per-record bins that it retrains when they fill [17], and LOFT, a new lock-free learned index, retrains in background threads and logs concurrent updates for replay [24]. Its authors note that many ALEX-OL data nodes retrain at nearly the same time. We trace it to bulk loading, predict it, and reduce its cost in ALEX-OL. We did not evaluate LOFT, which changes locking and retraining together.

***Tail latency from background work.*** Compaction in log-structured merge (LSM) trees [26] also causes latency spikes: writes stall when compaction falls behind [20]. SILK schedules flushes and compactions by I/O priority to reduce client tail latency [3]. Those spikes come from interference over shared resources. Ours come from operations blocked at a node being rebuilt, at predictable positions.

## 9 Conclusion

Each updatable learned index pays for concurrent inserts somewhere the standard benchmarks do not look: in synchronized bursts, in memory, in run time, or in lost keys. In gapped-array indexes, bulk loading makes the data nodes rebuild together at positions three source constants predict, where each rebuild blocks every operation that reaches its node. A buffer that exists only during a rebuild, together with background rebuilds, removes most of this cost without changing node size. We obtained every result with three standard systems practices: predicting where an index restructures before measuring it, timing every operation, and reading back every key. These practices cost little and apply to any concurrent index. The benchmark driver, the patches, and the results of all runs are public at https://github.com/mpi-dsg/GRE_Longitudinal/tree/wave-benchmark/wave-benchmark.

## Use of AI Tools

We used AI assistants, Anthropic's Claude and OpenAI's Codex, to edit the text, to write and run analysis and experiment scripts, and for literature research. The authors reviewed this work and take full responsibility for the contents of the paper.